\documentclass[fleqn,12pt]{wlscirep}
\usepackage{float}
\usepackage{setspace}
\usepackage{gensymb}
\usepackage{upgreek}
\usepackage{siunitx}
\usepackage{caption}
\usepackage[utf8]{inputenc}
\usepackage[english]{babel}
\DeclareCaptionLabelFormat{adja-page}{\hrulefill\\#1 #2 \emph{(previous page)}}
\title{A nanofabricated, monolithic, path-separated electron interferometer}

\author[1,+]{Akshay Agarwal}
\author[1,+]{Chung-Soo Kim}
\author[1]{Richard Hobbs}
\author[2,]{Dirk van Dyck}
\author[1,*]{Karl K. Berggren}
\affil[1]{Department of Electrical Engineering and Computer Science, Massachusetts Institute of Technology, Cambridge, MA USA}
\affil[2]{EMAT, University of Antwerp, Groenenborgerlaan 171 2020 Antwerp, Belgium}

\affil[*]{berggren@mit.edu}

\affil[+]{these authors contributed equally to this work}

\begin{abstract}
We report a self-aligned, monolithic electron interferometer, consisting of two 45 nm thick silicon layers separated by 20 $\upmu$m. This interferometer was fabricated from a single crystal silicon cantilever on a transmission electron microscope grid by gallium focused ion-beam milling. Using this interferometer, we demonstrate beam path-separation, and obtain interference fringes in a Mach-Zehnder geometry. The fringes have a period of 0.32 nm, which corresponds to the $\left[\bar{1}\bar{1}1\right]$ lattice planes of silicon, and a maximum contrast of 15\%, in an unmodified 200 kV transmission electron microscope. This design can potentially be scaled to millimeter-scale, and used in electron holography. It can also be applied to perform fundamental physics experiments, such as interaction-free measurement with electrons.    
\end{abstract}
\begin{document}
\flushbottom
\maketitle
\thispagestyle{empty}
\section*{Introduction}
Electron interferometers have been used in many applications such as demonstration of double-slit interference\cite{Merli1976} and the Ahronov-Bohm effect\cite{Tonomura1986}, exit-wave reconstruction \cite{Lichte1995}, and imaging magnetotactic bacteria \cite{DB1998}. Most of these applications were made possible by M\"ollenstedt and D\"uker's invention of the electron biprism in 1956 \cite{Mollenstedt1956}, which enabled wavefront-division interferometry in the electron microscope. This type of interferometry is fundamentally limited by the requirement of a highly coherent, field-emission electron source\cite{Cowley1992}; thermionic emission sources (such as LaB$_6$ and tungsten) lead to poor visibility of interference fringes from wavefront-division interferometry. This limitation also applies to the recently demonstrated double-slit electron interferometers \cite{Frabboni2007,Frabboni2008, Frabboni2011,Frabboni2012,Frabboni2015, Pope2013} that used focused-ion beam (FIB) fabricated slits. A second issue is that the integration of a biprism into a microscope requires considerable modification of the electron optical column.

The limitations of wavefront-division interferometry can potentially be overcome with an amplitude-division interferometer. Such an interferometer can provide much better interference fringe visibility (at the cost of reduced resolution \cite{Ru1995}) with low-coherence electron sources and is hence very useful for applications where the sensitivity of the measurement is important. An amplitude-division interferometer for electrons was first proposed and demonstrated by Marton and co-workers \cite{Marton1952,Marton1953,Marton1954}. This interferometer used three 10 nm thick, polycrystalline, epitaxially grown copper membranes that act as diffraction gratings to split and recombine the electron beam.  Multilayer interferometers using two layers at the edges of silicon crystals were later used by Dowell and Goodman \cite{Dowell1973,Dowell1977}, Buxton\cite{Buxton1978}, Rackham\cite{Rackham1977}, and Zhou \cite{Zhou1995,Zhou2001}. Designs by Matteucci\cite{Matteucci1981,Matteucci1982} and Ru\cite{Ru1993,Ru1995} also demonstrated interferometry with a thermionic source, and without significant modification the electron column optics, respectively. A combination of crystalline gratings and biprisms was also employed in interferometry and holography setups by Herring \cite{Herring1993, Herring1995}, and Mertens \cite{Mertens1999}. Besides crystalline gratings, electron diffraction nanofabricated gratings\cite{Jonsson1961} has also been used in amplitude-division interferometry. %Ito et al. used a scanning transmission electron microscope to fabricate 2D diffraction gratings \cite{Ito1993} and Fresnel lenses\cite{Ito1998}. Verbeek et al. \cite{Verbeeck2010}  and McMorran et al. \cite{McMorran2011} fabricated phase plates using FIB to create electron vortex beams. 
For example, Gronniger et al. \cite{Gronniger2005,Gronniger2006} and Bach et al. \cite{Bach2013} constructed Mach-Zehnder and Talbot-Lau electron interferometers with thermionic electron guns, using three large-area gratings fabricated by optical interference lithography.

Despite these advances, amplitude-division electron interferometers have not been widely adopted. This lack of utilization is primarily due to the stringent requirements of positioning and orientation for precise alignment of the interferometer, which have resulted in considerable modification of the electron column in previous efforts, just as for the biprism. For example,  Marton\cite{Marton1954} had to develop a  mechanical manipulator to control translation and rotation of each grating for alignment. Gronniger's experiment\cite{Gronniger2006} incorporated a laser interferometer to rotationally align the gratings to an accuracy of 1 mrad. Buxton and Zhou's double crystal interferometer overcame the requirement of alignment by using two silicon layers from the same crystal. However, it had limited applicability due to the small ( $\sim$ \SI{1}{\micro\metre}) gap between the crystals which made separation of interferometer paths difficult \cite{Missiroli1981}. Complete path separation is important to ensure that one of the beams can be manipulated without affecting the other.

In this work, we fabricated a self-aligned electron interferometer using FIB sculpting of a thick single crystal of silicon (110). We also demonstrated interferometry in the Mach-Zehnder configuration by directly imaging the interference between two electron beams diffracted from the gratings in a transmission electron microscope (TEM). The interferometer was integrated in the TEM with no modification of the electron column. Diffraction and interference experiments confirmed that our grating architecture was aligned to an accuracy of \SI{100}{\micro\radian}.
\begin{figure}
\includegraphics[width=1\textwidth]{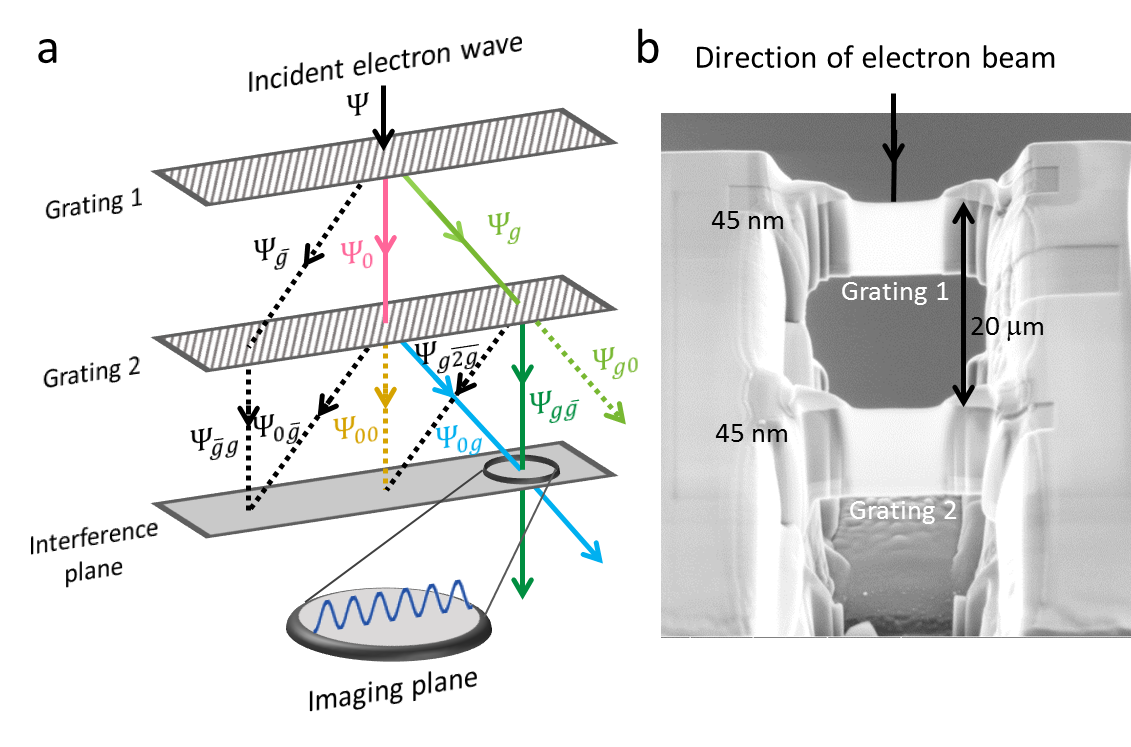}
\centering
\caption{Two-grating electron interferometer. (a) Schematic of diffraction with 1-D gratings.  The incident wave $\Psi$ is diffracted into zero ($\Psi_{\textbf{0}}$) and first ($\Psi_{\textbf{g}}$ and $\Psi_{\bar{\textbf{g}}}$) order beams by grating 1, each of which diffracts again from grating 2. Any two diffracted partial waves with $\Delta \textbf{g}_{\textrm{net}}\hspace{1pt}=\textbf{g}$ (\textrm{see text for definition}), such as $\Psi_{\textbf{0g}}$ and $\Psi_{\textbf{g}\bar{\textbf{g}}}$ interfere in the `interference plane', thereby imaging the lattice planes corresponding to \textbf{g} onto this plane.We placed a CCD camera conjugate to the interference plane to capture the interference pattern.(b) Scanning electron micrograph of two-grating interferometer with 20 $\pm$ \SI{0.1}{\micro\metre} gap between the gratings. The thickness of each grating is 45 $\pm$ 5 nm. This image was taken at a 52\degree tilt at an electron energy of 5 keV and working distance of 4 mm.}
\label{Fig1}
\end{figure}

Figure \ref{Fig1}(a) schematically depicts diffraction from the two-grating interferometer. To simplify the description we use a 1-D grating with lattice constant $a$ and depict only the zeroth and first order diffracted waves from each grating. $D$ denotes the gap between the two gratings. Solid lines represent the waves of interest in the interferometer. We use Zhou's notation \cite{Zhou2001} to denote the diffracted waves from the two gratings. The first grating splits the incident wave $\Psi$ into the zero ($\Psi_{\textbf{0}}$, pink) and first order ($\Psi_{\textbf{g}}$, light green and $\Psi_{\bar{\textbf{g}}}$, black) diffracted partial waves. Here $|\textbf{g}|=2\pi/a$ is the magnitude of the 1-D reciprocal lattice vector. Each of these waves is incident at a Bragg angle on the second grating and gets diffracted again, provided the two gratings are mutually aligned. The re-diffracted partial waves arising from $\Psi_{\textbf{0}}$ are $\Psi_{\textbf{00}}$ (yellow), $\Psi_{\textbf{0g}}$ (blue), and $\Psi_{\textbf{0}\bar{\textbf{g}}}$ (black), and similarly for $\Psi_{\textbf{g}}$ ($\Psi_{\textbf{g0}}$ (light green), $\Psi_{\textbf{g}\bar{\textbf{g}}}$ (dark green)) and $\Psi_{\bar{\textbf{g}}}$ ($\Psi_{\bar{\textbf{g}}\textbf{0}}$ and $\Psi_{\bar{\textbf{g}}\textbf{g}}$, both black). Defining $\textbf{g}_{\textrm{net}}$ as the sum of the subscript $\textbf{g}$-vectors for each wave, we see that any two diffracted waves $\Psi_{\textbf{g1g2}}$ and $\Psi_{\textbf{g3g4}}$ for which $|\Delta \textbf{g}_{\textrm{net}}|=|(\textbf{g3}+\textbf{g4})-(\textbf{g1}+\textbf{g2})| =0$, such as $\Psi_{\bar{\textbf{g}}\textbf{g}}$  and $\Psi_{\textbf{00}}$, or $\Psi_{\textbf{0}{\textbf{g}}}$ and $\Psi_{\textbf{g}0}$, emerge parallel to each other after diffraction from both gratings. Waves with $|\Delta \textbf{g}_{\textrm{net}}|=\textbf{g}$ such as $\Psi_{\textbf{0g}}$ and $\Psi_{\textbf{g}\bar{\textbf{g}}}$ overlap and interfere with each other. This interference occurs in a plane parallel to the two gratings and located $D$ units below the second grating. We will henceforth refer to this plane as the `interference plane'. In our experiments, we used $\Psi_{\textbf{0g}}$ and $\Psi_{\textbf{g}\bar{\textbf{g}}}$ to construct a separate-path interferometer in the TEM. An equivalent interferometer is formed by $\Psi_{\bar{\textbf{g}}\textbf{g}}$ and $\Psi_{\textbf{0}\bar{\textbf{g}}}$. The interference fringes can be read out by placing a third grating in the interference plane and recording the electron counts on an integrating detector positioned in the path of either of the output waves ($\Psi_{\textbf{0g}}$ and $\Psi_{\textbf{g}\bar{\textbf{g}}}$). Translation of the third grating perpendicular to the optical axis leads to oscillations in these counts due to change in the relative phase between the two interfering waves. Working in a TEM allowed us to observe interference fringes by directly imaging the interference plane, which precluded the need for a third grating.
\section*{Experiment}
Figure \ref{Fig1}(b) shows the fabricated two-grating structure that we used for interference experiments. As described in Methods, we fabricated our gratings on a workpiece consisting of single-crystalline silicon cantilevers suspended from a tungsten support grid using FIB milling (FEI Helios Nanolab 600 and 650). The grid could be inserted into a regular TEM sample holder. The gap between the gratings is 20 $\pm$ \SI{0.1}{\micro\metre} and the thickness of each grating in the region used in our experiments is 45 $\pm$ 5 nm. Another two-grating structure with \SI{2.5}{\micro\metre} gap between the gratings was used to characterize the alignment and coherence of diffracted beams as described in Discussion. The grid was mounted into the sample stage of a TEM (JEOL 2010F) for electron diffraction and interferometry experiments. These experiments were performed at an electron energy of 200 keV. 

As discussed previously, separating the paths of the interfering beams is critical to independently manipulating the phase of each beam. In order to determine the beam diameter, semi-angle of convergence $\alpha$, and grating separation for a separate path interferometer, we simulated the diffraction of 200 keV electrons from two gratings using the Gaussian-Schell model (GSM) \cite{Freiberg1982,McMorran2008,McMorran2008_2}. GSM assumes that the incident beam consists of a distribution of independent Gaussian modes and allows for the description of partially coherent beams using the mathematics of Gaussian beams. We used McMorran and Cronin's results on the diffraction of a GSM beam from two gratings\cite{McMorran2008_2}, with a beam spatial coherence length equal to 20\% of the beam diameter, in our simulations. This estimate of the spatial coherence was based on theoretical calculations for small condenser apertures \cite{Dronyak2009,Morishita2013}, and supported by preliminary experiments (described in Discussion) to characterize the beam coherence. We assumed that the degree of temporal coherence of the beam was close to 1, and hence ignored the effects of partial temporal coherence in our simulations \cite{WilliamsCartercomp}. We used the (000) and ($\bar{1}\bar{1}1$) diffracted beams of silicon to design our interferometer. Therefore, each grating in the simulation was one-dimensional with a period of 0.32 nm, which is equal to the period of the $\left[\bar{1}\bar{1}1\right]$ lattice planes. An important \textit{caveat} here is that the gratings in our simulation were amplitude gratings, while thin layers of silicon behave as phase gratings at the electron energies used in the TEM. However, this difference did not affect the diffraction angles, and hence the set of parameters that allowed the beams to separate, which was the primary focus of the simulations. We chose a beam diameter of 240 nm at the first grating and $\alpha=$ 4 mrad, with beam crossover (\textit{i.e.} beam focus) between the second grating and interference plane. With these parameters, the beam diameter in the the interference plane was 80 nm. The chosen parameters prevented overlap between the diffracted beams $\Psi_{\textbf{0}}$ and $\Psi_{\textbf{g}}$ on the second grating, and $\Psi_{\textbf{00}}$ and $\Psi_{\textbf{g}\bar{\textbf{g}}}$ at the interference plane, and thus ensured complete path separation. The choice of beam parameters was dictated by experimental considerations, as explained in the supplementary information. 
%With a beam diameter of 200-400 nm, our simulations indicated that a minimum gap of 12-\SI{25}{\micro\metre} between the gratings was required to avoid undesirable beam overlap on the second grating and the interference plane. We chose a gap of \SI{20}{\micro\metre} as this was the maximum separation possible on our TEM grids. Figure S4 (supplementary information) shows the allowed range of beam semi-convergence angles and diameters for $D=$ \SI{20}{\micro\metre}. We chose a beam diameter of 240 nm at the first grating and semi-angle of convergence of $\sim$ 4 mrad, with beam crossover (\textit{i.e.} beam focus) between the second grating and interference plane. With these parameters, the beam diameter in the plane of the second grating and the interference plane will be 80 nm. The choice of beam parameters was dictated by experimental considerations. A beam with smaller diameter and $\alpha$ than those used specified above required the use of a very small (\SI{10}{\micro\metre}) condenser aperture which reduced the intensity of the beams and thus increased the exposure time required to record interference fringes (using the procedure described below) . This resulted in poor fringe contrast due to stage vibrations. 

Figure \ref{Fig2}(a) shows the simulated diffraction of a GSM beam with these parameters from two 0.32-nm-period gratings separated by \SI{20}{\micro\metre}. The simulation included diffracted beams up to second order. For the following simulation and experimental results in this section, $\textbf{g}$ = ($\bar{1}\bar{1}1$). As described earlier, any two diffracted beams with $|\Delta \textbf{g}_{\textrm{net}}|=\textbf{g}$ overlap  in the interference plane, which for our interferometer was \SI{20} {\micro\metre} below the second grating. In figure \ref{Fig2}(b), we magnify the region around the interference plane to see the overlapping beams. Note that the fringe contrast in this image was caused by undersampling and consequent aliasing of the underlying lattice-spaced interference pattern in the simulation. However, the extent of these aliased fringes along the optical axis, $\Delta z$ $\sim$ \SI{2.7}{\micro\metre},  was the same as that of the actual interference fringes. $\Delta z$  is proportional to the spatial coherence of the beams, as discussed later. Figure \ref{Fig2}(c) shows a cross-section of the overlapping beams in the interference plane with further magnification and finer meshing; we obtained fringes with the period of the corresponding lattice, \textit{i.e.}, 0.32 nm. %There is also overlap between beams with $|\Delta \textbf{g}_{\textrm{net}}|=2\textbf{g}$, such as $\Psi_{\textbf{0 2g}}$ and $\Psi_{\textbf{g}\bar{\textbf{g}}}$, midway between the second grating and the interference plane. However, the fringes formed from this interference will have half the period of the corresponding lattice. For $\textbf{g}$=($\bar{1}\bar{1}1$) this period is equal to 0.16 nm which is very close to the resolution limit of our TEM. We therefore focus only on fringes in the interference plane. 
\begin{figure}
\centering
\captionsetup{labelformat=empty}
\includegraphics[width=0.8\textwidth]{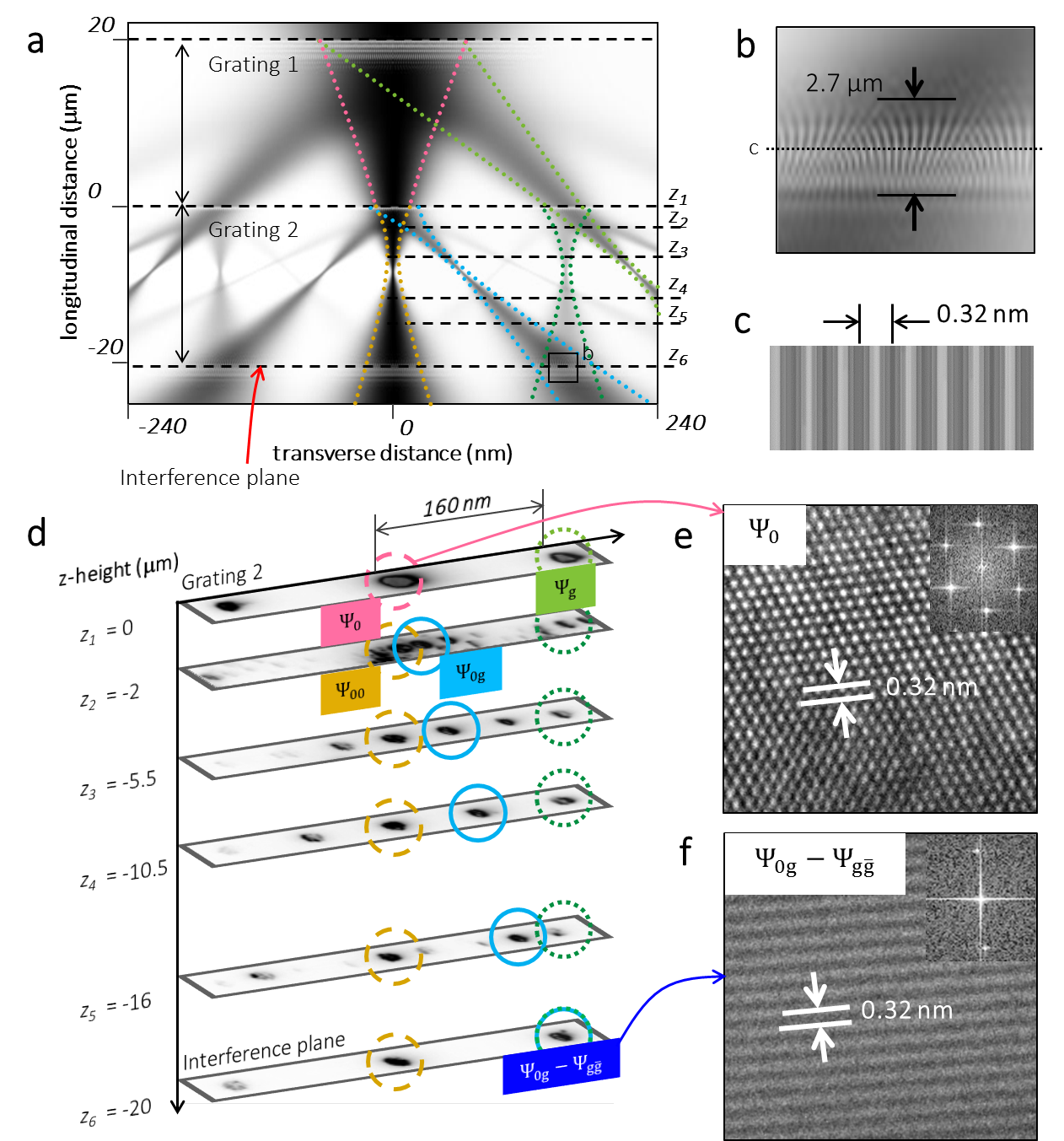}
\caption{}
\label{Fig2}
\end{figure}
%\clearpage
\begin{figure}
\captionsetup{labelformat=adja-page}  %This adds 'previous page'
    \ContinuedFloat
    \caption{Electron interferometry with the two-grating structure.(a) GSM simulation of diffraction from the \SI{20}{\micro\metre} interferometer. The black dashed lines represent imaging planes at different stage heights $z$, as explained in the text. In the interference plane, the two beams $\Psi_{\textbf{0g}}$ (blue) and $\Psi_{\textbf{g}\bar{\textbf{g}}}$ (dark green) overlap. (b) Magnified view of the region around the interference plane as indicated in (a). For a spatial coherence length equal to 20 \% of the beam diameter, the interference fringes extend for $\Delta z \sim$ \SI{2.7}{\micro\metre} along the optical axis. The contrast seen in this image is caused by undersampling of the actual interference fringes, as explained in the text. (c) Magnified cross-section of the overlapping beams at the interference plane, as indicated in (b), showing fringes with the periodicity of the $\left[\bar{1}\bar{1}1\right]$ planes, 0.32 nm. (d) ($z_1$ = 0) $\Psi_{\textbf{0}}$ (center, pink circle), $\Psi_{\textbf{g}}$ (right, light green circle) and $\Psi_{\bar{\textbf{g}}}$ (left) diffracted beams on the second grating. ($z_2$ = -\SI{2}{\micro\metre}) $\Psi_{\textbf{0g}}$ (circled in blue) separates out from $\Psi_{\textbf{00}}$ (circled in yellow). $\Psi_{\textbf{g}\bar{\textbf{g}}}$ is circled in dark green on the right. ($z_3$ = -\SI{5.5}{\micro\metre} , $z_4$ =-\SI{10.5}{\micro\metre} and $z_5$ = -\SI{16}{\micro\metre}) $\Psi_{\textbf{0g}}$ moves towards  $\Psi_{\textbf{g}\bar{\textbf{g}}}$. The measured distances between the beams are included in the text. ($z_6$ = - \SI{20}{\micro\metre}) The two beams $\Psi_{\textbf{0g}}$ and $\Psi_{\textbf{g}\bar{\textbf{g}}}$ overlap and interfere.(e) TEM micrograph of the lattice of the second grating from the $\Psi_{\textbf{0}}$ beam at $z_1$=0. Inset shows the Fourier transform of the image, with multiple spots corresponding to the different lattice planes of silicon (110). (f) TEM micrograph of fringes from the interference of  $\Psi_{\textbf{0g}}$ and $\Psi_{\textbf{g}\bar{\textbf{g}}}$ beams at $z_6$ = -\SI{20}{\micro\metre} with a period 0.32 nm. The inset shows the Fourier transform of the image. Only one set of points (corresponding to $\textbf{g}$=($\bar{1}\bar{1}1$)) are seen around the central spot, confirming the origin of the fringes.}
\label{Fig2}
\end{figure}

In Figure \ref{Fig2}(d), we show experimental demonstration of this interferometer in the Mach-Zehnder geometry. Note that we focused the electron beams very close to the image plane, to obtain the images in this figure and make the movement of various beams easier to follow. However, to get high-resolution lattice/interference fringe images, we defocused the beams to the designed diameter and $\alpha$. 

In order to see the various beams diffracted by the two gratings, we translated the interferometer vertically by changing the TEM stage height $z$. This enabled us to successively image planes between second grating and interference plane, and thus follow the evolution of the diffracted beams in these planes. We started our experiment with the second (lower) grating in the eucentric plane. In figure \ref{Fig2}(d) we denote the stage height here as $z_1$ = \SI{0}{\micro\metre}. At this height, we imaged the primary and first order diffracted beams ($\Psi_{\textbf{0}}$ and $\Psi_{\textbf{g}}$, circled in pink and light green, respectively) on this grating. The separation $s$ between $\Psi_{\textbf{0}}$ and $\Psi_{\textbf{g}}$ was 160 nm as expected for $D$ = \SI{20}{\micro\metre} ($s=2\theta_{B}D$ where $2\theta_{B} \simeq \lambda_{\textrm{electron}}/a$). As seen in figure \ref{Fig2}(e), upon underfocusing the beams to the designed beam diameter (80 nm) at the second grating and imaging $\Psi_{\textbf{0}}$ at high-resolution, we obtained a lattice-resolved image of the crystalline silicon. 
We then translated the stage to $z_2$ = \SI{2}{\micro\metre} below the second grating, to image the beams diffracted by this grating. $\Psi_{\textbf{0g}}$ (circled in blue) was visible at a distance of 15 nm from $\Psi_{\textbf{00}}$. $\Psi_{\textbf{g}\bar{\textbf{g}}}$ (circled in dark green) was at a distance of 160 nm from $\Psi_{\textbf{00}}$. $z_3$ = \SI{5.5}{\micro\metre} below the second grating, the distance between $\Psi_{\textbf{0g}}$ and $\Psi_{\textbf{00}}$ increased to 42 nm. $\Psi_{\textbf{g}\bar{\textbf{g}}}$ (circled in dark green) remained 160 nm away from $\Psi_{\textbf{00}}$. $\Psi_{\textbf{g}\bar{\textbf{2g}}}$ (not circled) was also visible between $\Psi_{\textbf{0g}}$ and $\Psi_{\textbf{g}\bar{\textbf{g}}}$. On moving $z_4$ = 10.5 and $z_5$ = \SI{16}{\micro\metre} below the second grating, we observed that $\Psi_{\textbf{0g}}$ continued moving away from $\Psi_{\textbf{00}}$ and towards $\Psi_{\textbf{g}\bar{\textbf{g}}}$. The distance between $\Psi_{\textbf{0g}}$ and $\Psi_{\textbf{g}\bar{\textbf{g}}}$ was 78 and 29 nm for $z_4$ and $z_5$ respectively. Finally, when we reached $z_6$ = \SI{20}{\micro\metre} below the second grating, $\Psi_{\textbf{0g}}$ and $\Psi_{\textbf{g}\bar{\textbf{g}}}$  overlapped completely; the CCD camera was now conjugate to the interference plane. As shown in figure \ref{Fig2}(f), we observed interference fringes with a period of 0.32 nm within the overlap spot. We took this image by overfocusing the beam to a diameter of 80 nm, so that the beam diameter and $\alpha$ were at their designed values. Since the fringe contrast was quite low ( $<$20 \%), we used the Fourier transform of the live image (inset, figure \ref{Fig2}(f)) to monitor the appearance of the fringes. The presence of a single set of spots in the Fourier transform (corresponding to $\textbf{g}=(\bar{1}\bar{1}1$)) confirmed that these fringes were formed due to the interference between $\Psi_{\textbf{0g}}$ and $\Psi_{\textbf{g}\bar{\textbf{g}}}$. 
\section*{Discussion}
Successful demonstration of interference was critically dependent on the alignment between the two gratings. Further, the diffracted beams from each grating had to be sufficiently coherent to form visible fringes upon interference. Therefore, before performing electron interferometry experiments with the \SI{20}{\micro\metre} interferometer, we checked the alignment of our two-grating structures and the coherence of the diffracted beams by using parallel and convergent electron diffraction through a \SI{2.5}{\micro\metre}-gap structure. We also performed these tests for the \SI{20}{\micro\metre}-gap structure, as reported in the supplementary information. Although the tests indicated that the \SI{20}{\micro\metre}-gap structure is well-aligned, the convergent beam diffraction results were difficult to interpret due to limitations of our TEM. We discuss these limitations in detail later. 

\begin{figure}
\centering
\includegraphics[width=0.8\textwidth]{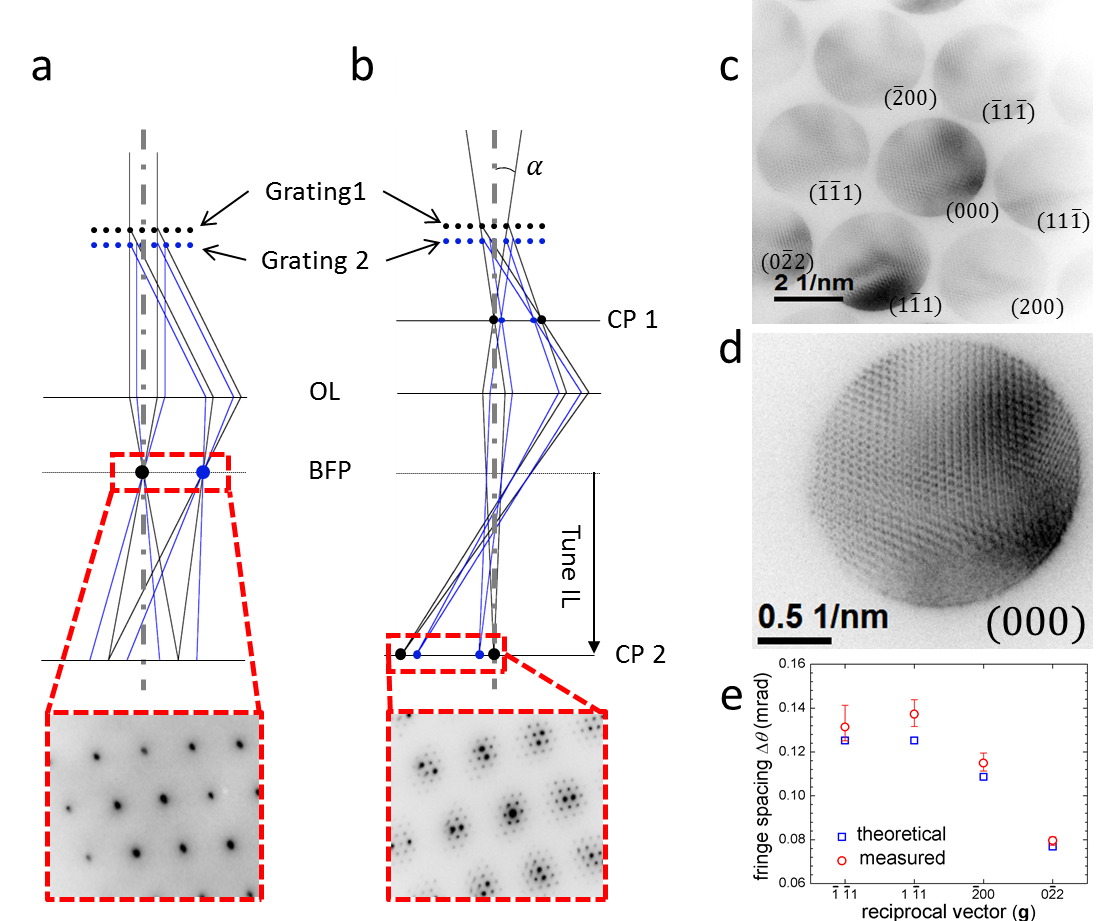} %need to rotate!!
\caption{Parallel and convergent-beam diffraction from the two-grating structure.(a) Ray diagram for a nearly parallel incident beam (small $\alpha$). Diffracted beams with $\Delta \textbf{g}_{\textrm{net}}(\textrm{see text for explanation})=0$ are focused to the same point in the back-focal plane (BFP) of the objective lens (OL). Thus the diffraction pattern is the same as single-layer silicon as seen in the experimentally obtained SADP (red box). The experimental SADP is for $\alpha$ = 0.1 mrad.  (b) Ray diagram for a convergent incident beam (large $\alpha$). The spots in the BFP broaden to disks formed by overlap between waves with $\Delta \textbf{g}_{\textrm{net}}=0$ which leads to interference fringes within each spot. At the first and second crossover plane (CP 1 and CP 2 respectively), these beams focus at horizontally displaced points due to the gap between the gratings. The red box is the experimental diffraction pattern in the second CP with $\alpha$=4 mrad, showing multiple closely spaced spots due to this horizontal displacement. (c) BFP diffraction pattern with $\alpha$=4 mrad showing interference fringes for the \SI{2.5} {\micro\metre}-gap sample. The $\textbf{g}_\textrm{net}$ corresponding to each spot is indicated. (d) Magnified view of the $\textbf{g}_\textrm{net}$ = (000) spot showing interference fringes (e) Angular separation of fringes in the (000) spot, \textit{i.e.}, formed by interference between diffraction orders with $\textbf{g}_{\textrm{net}}=0$. In this case, the $\textbf{g}$-vectors from the two gratings must be equal in magnitude and oppositely directed. These vectors are indicated on the \textit{x}-axis.}
\label{Fig3}
\end{figure}

For testing alignment, we took a selected area diffraction pattern (SADP) with a wide, nearly parallel electron beam ($\alpha=$ 0.2 mrad); figure \ref{Fig3}(a) shows a ray diagram for this situation. After diffraction from the two gratings, waves a common value of $\textbf{g}_{\textrm{net}}$, \textit{i.e.},  with $\Delta \textbf{g}_{\textrm{net}} =0$ are parallel to each other. Therefore, we expect these waves to be focused at the same point at the back focal plane (BFP) of the TEM objective lens. Hence, the focused SADP should be the same as for single-layer silicon, provided the two gratings are well-aligned. This prediction was confirmed in the experimentally observed SADP, as shown in the box at the bottom of figure \ref{Fig3}(a). We did not observe any displacement between the focused diffraction spots from the two gratings for camera lengths up to 200 cm.

For testing coherence between beams diffracted from the two gratings, we increased $\alpha$ to 4 mrad; in figure \ref{Fig3}(b) we depict the ray diagram for electron diffraction with a convergent beam. As a result of the beam convergence the focused spots in the BFP broadened into disks, with each disk formed by overlap between beams with $\Delta \textbf{g}_{\textrm{net}} =0$. We focused the diffraction pattern by tuning the intermediate lens (IL) current. This change in the IL current changed the plane being imaged from the BFP to the second `crossover plane' (CP 2), which was an image of the focused beams at the first crossover plane (CP 1) below the two gratings. In this plane, the gap between the two gratings led to horizontal displacement between the focused spots from beams with $\Delta \textbf{g}_{\textrm{net}} =0$ \cite {Mertens1999_2}. The box at the bottom of figure \ref{Fig3}(b) shows the experimental SADP with for $\alpha =4$ mrad. This SADP is reminiscent of a Moir\'e pattern, except that the extra diffraction spots were created not due to different lattice constants\cite{WilliamsCarter}, but rather due to the gap between the gratings. Note that this displacement was unrelated to the misalignment-induced displacement at the BFP expected for a parallel beam. The supplementary information contains an extended discussion of convergent beam diffraction from the two-grating structures, along with supporting experiments to verify the mechanism outlined above.

Within each spot in the BFP, we observed interference fringes with multiple orientations and periods, as seen in the SADP in figure \ref{Fig3}(c). We will henceforth refer to these fringes as `BFP fringes' to differentiate them from the imaging plane fringes obtained with the \SI{20}{\micro\metre} interferometer. Interference between $\Psi_{\textbf{00}}$, $\Psi_{\textbf{g}\bar{\textbf{g}}}$ and $\Psi_{\bar{\textbf{g}}\textbf{g}}$ led to BFP fringes within the zero-order spot (since $\textbf{g}_{\textrm{net}}$ = 0 for each of these beams) perpendicular to $\textbf{g}$, as seen in figure \ref{Fig3}(d). Similarly, interference between $\Psi_{\textbf{0}\bar{\textbf{g}}}$ and $\Psi_{\textbf{g}\bar{\textbf{2g}}}$ resulted in BFP fringes in the $\textbf{g}_{\textrm{net}} = \bar{\textbf{g}}$ spot. Inclusion of all the silicon reciprocal lattice vectors in this description would lead to the different interference fringe orientations and periods in figures \ref{Fig3}(c) and (d). These BFP fringes confirmed that the beams diffracted from the first and second gratings were at least partially coherent with each other. In previous work by Buxton and Zhou the angular separation between these fringes was estimated to be $\Delta \theta \sim a/D$ \cite {Buxton1978, Zhou1995, Zhou2001}. Physically, a larger reciprocal lattice vector and/or gap between the gratings increases the angle of intersection between the overlapping beams in the BFP, thus reducing the period of the resulting fringes. Importantly, Buxton and Zhou's estimate for $\Delta \theta$ does not depend on $\alpha$. We measured $\Delta \theta$ for $\alpha = 4, 2, 0.9,$ and $0.5$ mrad, by varying the size of the selected-area diffraction (SAD) aperture, keeping all lens currents constant. In figure \ref{Fig3}(e) we compare the mean of $\Delta \theta$ for these values of $\alpha$ with  Buxton and Zhou's estimate, for fringes within the zero order spot with $\textbf{g}$ = ($\bar{1}\bar{1}1$), ($1\bar{1}1$), ($\bar{2}00$) and ($0\bar{2}2$). The experimental values agreed with the estimate to an accuracy of 5\%, 9.6\%, 5.7\% and 3.7\% for the four value of $\textbf{g}$ respectively. The variation in the difference between the experimental values and the estimate was due to residual astigmatism in the objective lens. The change in $\Delta \theta$ with $\alpha$ was smaller than 3\% of the mean for all orientations of $\textbf{g}$. Thus, the chief source of error in $\Delta \theta$ was the pixel size of our CCD detector. The error bars for each value of $\textbf{g}$ in figure \ref{Fig3}(e) indicate the range of $\Delta \theta$ with an error of one pixel. Further, the contrast of the BFP fringes increased from 15\% (for $\alpha$ = 4 mrad) to 33\% (for $\alpha$ = 0.5 mrad). As noted earlier, we expected the degree of temporal coherence to be close to 1. Hence, the fringe contrast can be used as an estimate of the degree of spatial coherence of the diffracted beams\cite{Lichte1995} for different SAD apertures. In the GSM interference simulations we used a slightly higher value of the degree of spatial coherence (20\%) than that measured here (15\%) for $\alpha$ = 4 mrad, because of contrast reduction due to unequal amplitudes of the interfering beams. We elaborate on this point later.

We obtained similar SADP from the \SI{20}{\micro\metre}-gap-structure with parallel and convergent beams, as discussed in the supplementary information. The BFP fringe period was expected to be $\sim$ 10 times smaller than that for the \SI{2.5}{\micro\metre} gap sample, which was very close to the resolution limit of the CCD detector of our TEM. We were thus unable to image the BFP fringes with the \SI{20}{\micro\metre}-gap-structure. While obtaining the imaging plane fringes with this structure, as described earlier, we noted a slight displacement between the focused spots from each wave, from which we estimated a misalignment of $\sim$ \SI{100}{\micro\radian} between the two gratings. We discuss possible causes for this misalignment later. However, note that this misalignment was an order of magnitude lower than both the maximum tolerance for Marton's interferometer (1.2 mrad) \cite{Marton1954, Simpson1954} and the misalignment for Gronniger's interferometer (1 mrad) \cite{Gronniger2006}. Also, as noted earlier, we were able to position each grating with an accuracy of $\Delta D =100$ nm. The fractional error in positioning of the gratings $\Delta D/D=0.005$ was greater than Marton's tolerance specifications ($\Delta D/D=0.004$) and a factor of $\sim$ 6 larger than the corresponding value for Gronniger. More careful fabrication should allow us to position our gratings with better accuracy.

We again stress that the BFP fringes obtained here are different in origin from the imaging plane fringes reported with the \SI{20}{\micro\metre} interferometer. In the former case, beams with $\Delta \textbf{g}_{\textrm{net}} =0$ interfere in the BFP due to focusing by the objective lens, while in the latter case, beams with $\Delta \textbf{g}_{\textrm{net}} =$ \textbf{g} interfere due to diffraction from the second grating. Although the BFP fringes are useful for characterizing the coherence of the diffracted beams, the small gap results in the beams not being fully separated on the second grating. Hence this structure cannot be used in experiments that require the placement of a sample in the path of one of the beams, \textit{i.e.}, as a path-separated interferometer.
%\clearpage

We now return to the imaging plane fringes with the \SI{20}{\micro\metre} interferometer. The observed fringe spacing of 0.32 nm would have also been produced by Talbot self-imaging\cite{McMorran2009,Cronin2009}. However, the separation between the two gratings was $\sim$ 250 times the Talbot length $z_T$ for the ($\bar{1}\bar{1}1$) lattice planes of silicon ($z_T=2a^2/\lambda_{\textrm{electron}}$ = 82 nm for 200 keV electrons), which made Talbot fringes unlikely. A Moir\'e deflectometer\cite{Oberthaler1996} would have also produced fringes of the same period. The direct imaging method employed here, which showed the separation and overlap of diffracted beams, along with our measurement of beam coherence, made this explanation unlikely too. Thus, the observed fringes could be attributed to coherent overlap between the diffracted beams in a Mach-Zehnder geometry.

The fringe images were captured with an exposure time between 1 and 5 seconds. Longer exposures lead to blurring due to mechanical vibrations in the sample stage, while shorter exposures result in poor signal-to-noise ratio.

The spatial coherence length of the electron beam can be interpreted as the diameter of a disk of points, around any given point in the beam, that have a fixed mutual phase relationship. The wider this disk, the greater the extent of coherent interference along the optical axis. This increases $\Delta z$, the distance along the optical axis over which the fringes persist.  In our experiment, $\Delta z \sim$ \SI{3}{\micro\metre}. This value is close to the estimate of \SI{2.7}{\micro\metre} from our simulations, as shown in figure \ref{Fig2}(b), which further supports our assumption of the spatial coherence length of the beam in the simulation.

The maximum fringe contrast observed was 15\%, similar to the contrast for the BFP fringes with $\alpha= 4$ mrad described earlier. For two interfering beams of equal amplitude the fringe contrast is ideally equal to the degree of coherence. However, as can be seen from the images at $z_4$ and $z_5$ in figure \ref{Fig2}(d), the intensities of the $\Psi_{\textbf{0g}}$ and $\Psi_{\textbf{g}\bar{\textbf{g}}}$ beams were quite different. The ratio of the average intensity of the two beams from the image at $z4$ was 0.38. This difference in intensity reduced the fringe contrast by a factor of $\sim$ 0.9 from its ideal value. Tilting to the two-beam condition is a possible solution to enhancing the intensities of the interfering beams and thus improving contrast; however the slight misalignment between the two gratings noted earlier was sufficient to prevent us from achieving the two-beam condition simultaneously for both crystals. This misalignment is expected to be due to small bending of the crystals during fabrication. Another possible source of misalignment is the rotation of the electrons in the objective lens pre-field in which the sample is immersed. 

Due to bending and variations in thickness in each of the two gratings, the relative intensities of the diffracted beams varied with translation in the plane of the gratings (the \textit{x-y} plane). Since translation along the optical axis ($z$) led to small translations in the \textit{x-y} plane too, the intensities of the beams changed as we moved from the plane of the second grating to the overlap plane. This can be seen in the reduction of the intensity of the $\Psi_{\bar{\textbf{g}}}$ spot between the images at $z_1$ and $z_2$ in figure \ref{Fig2}(d). 
 \begin{figure}
\centering
\includegraphics[width=0.8\textwidth]{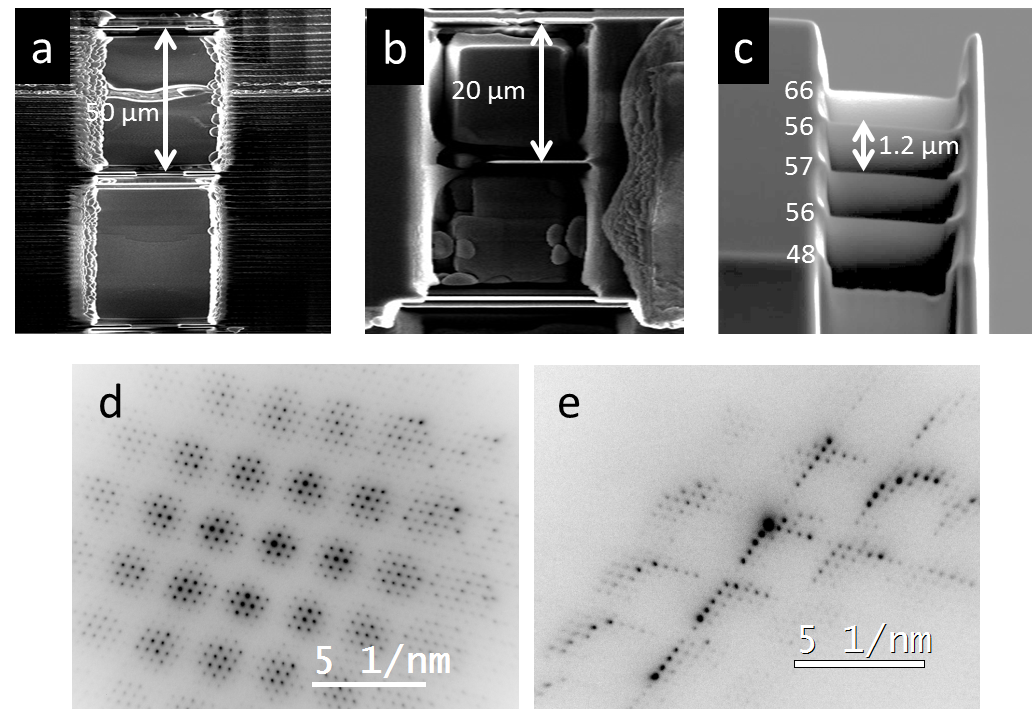}
\caption{ Controlling the geometry of the grating interferometers. Side-view SEM micrographs of three-grating structure with (a) 50 and (b) \SI{20}{\micro\metre} gap between the gratings, showing control over the gap and lateral area of each grating respectively. (c) 52\degree-tilt SEM micrograph of five grating structure with \SI{1.2}{\micro\metre} gap between the gratings, showing control over the number of gratings The thickness of each grating (in nm) is indicated. (d) Convergent-beam diffraction pattern from structure in (b). (e)  Convergent-beam diffraction pattern from structure in (c). }
\label{Fig4}
\end{figure}

This interferometer design can easily be scaled to larger gaps and numbers of gratings, which would facilitate its use in potential interferometry and holography setups by simplifying the placement of a sample and application of a field differential between the two beams. Figure \ref{Fig4}(a) shows a fabricated structure with \SI{50}{\micro\metre} gap between the gratings. We are currently limited by the thickness of commercially available TEM grids. Figure \ref{Fig4}(b) shows a three-grating structure with different lateral area of each grating, which allowed us to study diffraction through one, two or three gratings separately. Figure \ref{Fig4}(c) shows a five-grating structure with \SI{1.2}{\micro\metre} gap between the gratings, demonstrating control over the number of gratings in the fabricated structures. Figures \ref{Fig4}(d) and (e) show convergent beam diffraction from the grating structures in figures \ref{Fig4}(b) and (c) respectively, again showing multiple closely-spaced spots, as discussed for the two-grating structure.
 
Although we fabricated a three-grating interferometer (as shown in figure \ref{Fig4}(c)) to attempt a Marton-type experiment\cite{Marton1954}, the interpretation of this experiment was hindered by contrast fluctuations. These fluctuations were again caused by bending and thickness variation in each grating. In addition to interference effects, the intensity of each diffracted beam was also affected by the thickness of each grating. The difficulty in determining the exact thickness at every point of each grating made it challenging to separate this effect from the interference effects. This issue can be addressed by fabricating very thin ($\sim$ 10 nm) gratings to suppress dynamical diffraction effects. In the two-grating results described earlier, this problem was circumvented by effectively replacing the third grating with a screen on which the interference was imaged. 
\section*{Conclusion}
We have fabricated a monolithic, two-grating electron interferometer, which showed a misalignment of $\sim$ \SI{100}{\micro\radian}. This was an order of magnitude lower than similar designs reported previously. We demonstrated a path-separated electron interferometer in the Mach-Zehnder geometry, and obtained interference fringes with a period of 0.32 nm. The fringe contrast was used to estimate the spatial coherence of the TEM electron beam to be $\sim$ 20 \%.

This interferometer design is self-aligned, configurable, scalable to larger dimensions, and continues progress towards electron interferometry and holography in a conventional TEM with no modification of the optical column or sample holder \cite{Ru1995,Matteucci1981}. It could also be incorporated into a specially-designed electron-optic column for specific applications. The separation of paths on the second grating makes it feasible to place an absorbing object in the path of one of the beams, which may allow the implementation, with electrons, of Elitzur and Vaidman’s scheme for interaction-free imaging \cite{EV1993,Kruit2016}. In the same vein, we can also configure the gratings in order to implement multiple and repeated quantum interrogation of distinct absorbing objects\cite{Kwiat1995,Wang2016}. An important requirement for such structures is that the error in positioning of each grating ($\sim$ 100 nm as noted earlier) be smaller than $\Delta z$, to ensure coherent interference on each grating. This requirement is met by our design.  A major challenge that will need to be addressed is the variations in thickness of each grating which would make interpretation of any which-path experiment difficult.
\section*{Methods}
\subsection*{FIB Fabrication}
We fabricated the two-grating interferometers by gallium FIB milling (FEI Helios Nanolab 600 and 650) of single-crystal silicon (110) cantilevers on tungsten TEM grids (Nanomesh, from Hitachi High-Tech). The gratings were made on one monolithic silicon (110) crystal cantilever with two thicknesses (5 and \SI{40} {\micro\metre}). Figure S1 in the supplementary information summarizes the steps in our fabrication process. We initially placed the cantilever perpendicular to the ion-beam optical axis. The first step was milling of windows through the 5/\SI{40}{\micro\metre} thick silicon cantilevers using 30 kV gallium ion beam. These windows defined the lateral extent of the gratings. They also acted as a milling stopper, and helped to reduce material re-deposition and secondary sputtering in subsequent steps. We then placed the cantilever along the direction of the optical axis and deposited two \SI{3}{\micro\metre} thick platinum layers to define the gratings, and protect them from ion-beam damage. Next we milled the unprotected silicon at 30 kV and 21 nA beam current. This step at large current and energy milled most of the silicon between the gratings. It was important here to leave substantial ($\sim$ 500 nm) thickness at each grating to allow for some beam focusing errors and resulting damage in the non-milled area. The gratings were then polished, first at successively lower currents (down to 50 pA) and then at lower energies (down to 2 kV) to give the final structure. The polishing step thinned down the gratings to $<$ 50 nm and removed most of the amorphous layer formed from ion-beam damage. The final polishing was done at a slight tilt (up to 5\degree in either direction) to improve the uniformity of thickness each grating. We restricted the lateral dimensions of each grating to be \SI{10}{\micro\metre} by \SI{10}{\micro\metre} to avoid bending of the membranes.
\section*{Acknowledgments}
The authors would like to acknowledge helpful discussions and feedback from Dr. Pieter Kruit, Dr. Vitor Manfrinato, Dr. Jo Verbeek, Dr. Ivan Lobato and Dr. Ricardo Egoavil. The authors would also like to thank Stijn van Den Broeck for FIB fabrication of the two-grating interferometers. Some of the experiments were conducted using microscopy facilities at EMAT, University of Antwerp. This work also made use of the Shared Experimental Facilities supported in part by the MRSEC Program of the National Science Foundation under award number DMR-1419807. This work was supported by the Gordon and Betty Moore Foundation, and the National Research Foundation of Korea (NRF) grant funded by the Korea government(MEST) (No. 2013R1A6A3A03065200).
\section*{Author contributions statement}
A.A., C.S.K., R.H., and K.K.B. conceived the experiments; A.A. and C.S.K. conducted the experiments. D.V.D. developed the mechanism for electron diffraction from multi-gratings. All authors analyzed the results and reviewed the manuscript. 

\section*{Additional information}
The authors declare no competing financial interests.

\bibliography{ref.bib}

\end{document}

\maketitle
\section{Fabrication}
\begin{figure}
\includegraphics[width=1\textwidth]{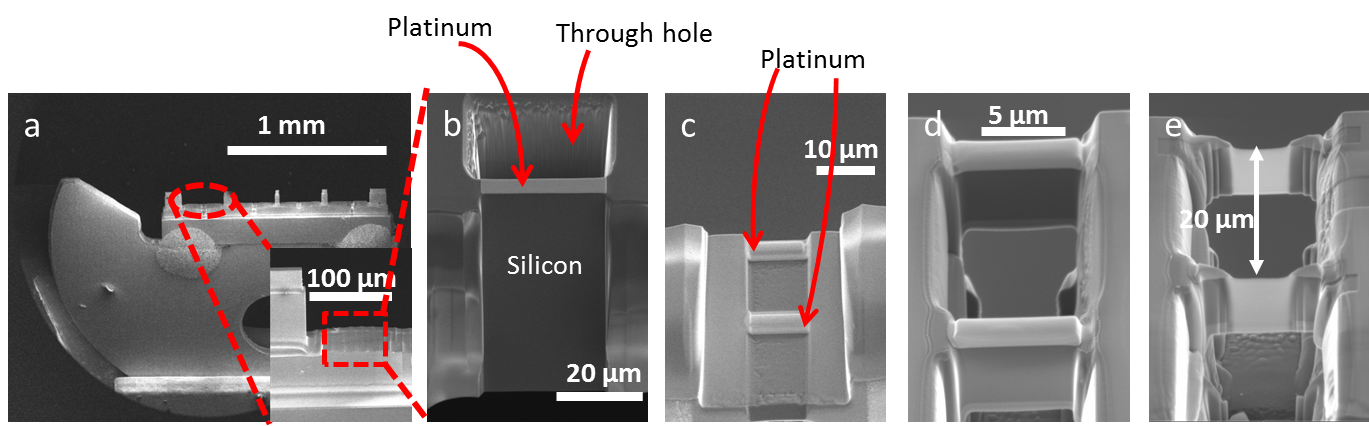}
\caption{FIB fabrication of the monolithic two-grating interferometer. (a) We started with single-crystal silicon cantilevers on a regular TEM grid. Inset shows a zoom of one of the cantilevers. (b) We deposited platinum on the top surface and milled a window through the cantilever to define the lateral area of the gratings (c) The grid was rotated by 90\degree and two \SI{3}{\micro\metre}- thick platinum layers were deposited on the silicon to protect the gratings (d) The silicon between the platinum layers was milled at high ion-beam energy (30 kV) and current (21 pA) (e) The two gratings were thinned down and polished to a final thickness of $\sim$ 45 nm by lowering the ion-beam energy down to 2 kV. }
\label{fab}
\end{figure}

The fabrication process, as described in the Methods section, is outlined in figure S\ref{fab}.
\section{Mechanism of electron diffraction from multiple crystal gratings}
\begin{figure}
\centering
\includegraphics[width=0.8\textwidth]{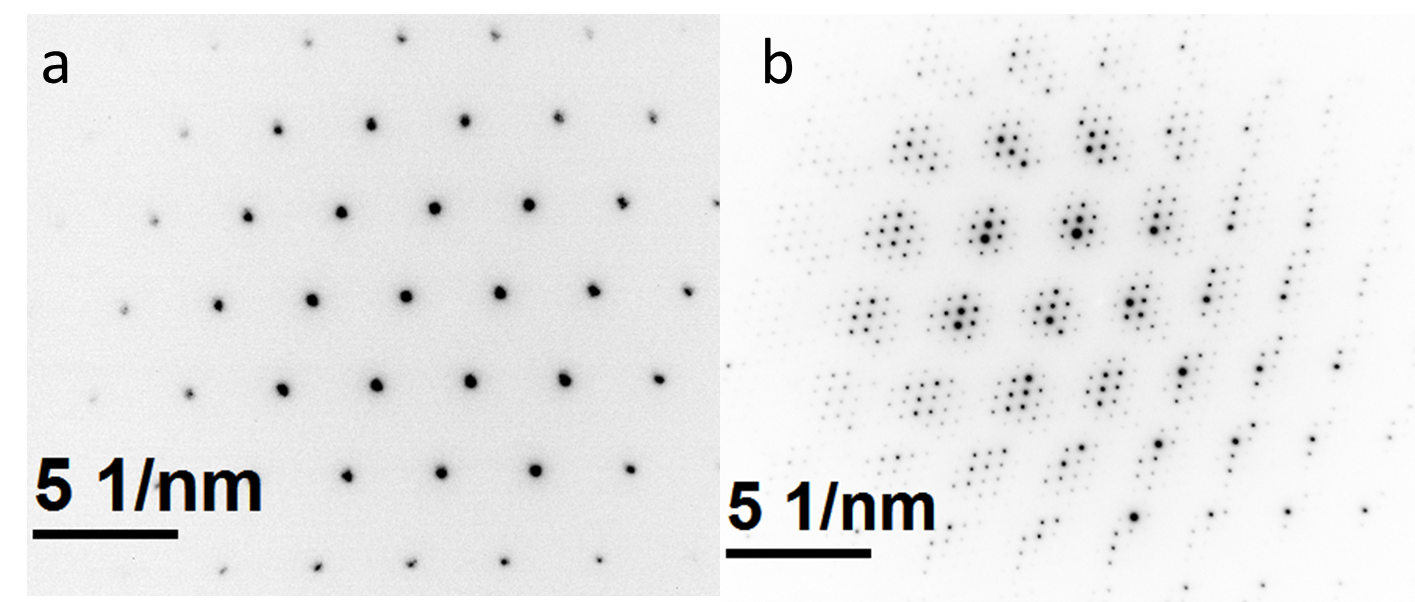}
\caption{Electron diffraction from \SI{20}{\micro\metre}-gap-structure. (a) With a nearly parallel beam, we obtained a diffraction identical to single layer silicon. (b) With a convergent beam we obtained multiple closely-spaced spots just as for the \SI{2.5}{\micro\metre}-gap-structure.}
\label{fig:20microndiff}
\end{figure}

We used electron diffraction from the two-grating structure to test the mutual alignment of the gratings as described in the paper. As shown in figure S\ref{fig:20microndiff}(a), the two-grating diffraction pattern from the \SI{20}{\micro\metre} structure with a nearly parallel (convergence semi-angle $\alpha=$ 0.2 mrad) beam is identical to that from the \SI{2.5}{\micro\metre}-gap-structure.

The convergent beam diffraction pattern in figure S\ref{fig:20microndiff}(b) can be considered as a regular silicon (110) DP on which another DP with the same symmetry but large demagnification is superimposed. The demagnification of the smaller DP in S\ref{fig:20microndiff}(b) is 0.15. Any attempt to explain this DP as a regular Moir\'e pattern runs into difficulties because a  $\sim$ 15 \% difference in lattice constant between the two gratings (induced by stress or thermal expansion) is unlikely. Further, the disappearance of the demagnified pattern for a nearly parallel beam also cannot be explained using this mechanism. This indicates that the DP is caused by the convergence of the beam.

\subsection{Convergent beam diffraction from a single-grating structure}
A convergent beam can be thought of as a set of plane waves from all the directions within the cone. When this beam is incident on a grating such that it is focused some distance below the grating, each incident plane is diffracted at the grating and generates plane waves that leave the crystal in the directions given by Bragg’s law. If the incident beam is inclined, the diffracted beams are inclined at the same angle. Thus each diffracted beam will leave the grating as a convergent cone but with the symmetry axis along the Bragg direction and will also converge to a small spot in the same focus plane below the sample as the incident beam, which we called the first crossover plane CP1 in figure 3 (b). The position of CP1 only depends on the setting of the condenser lens and not the position of the object. Crucially, the distance between the object and CP1 determines the gap between the spots in the CP1; the smaller this distance, the smaller the gap between the spots. 

\subsection{Convergent beam diffraction from a multi-grating structure}
With two (or more) gratings, the CP1 of the convergent beam remains the same, but now, due to the different vertical positions of the gratings along the optical axis, the distance between the focused diffracted spots from each grating is different. Therefore at CP1 we will see multiple superimposed DPs with the same symmetry but different magnifications. To image these focused beams, we need to tune the intermediate lens current to move below the back focal plane (where the convergent beams result in disks as seen in figure 3 (c) to the `second crossover plane' CP 2, as shown in figure 3(b). At this plane, we image the CP 1 and hence see the DP with multiple, superimposed magnifications. By taking all possible orders of diffraction into account, rather than just the first as in figure 3 (b), we can build up the experimental DP. This imaging mode bears some resemblance to conventional convergent beam electron diffraction (CBED), with the important difference that the beam crossover in CBED is focused inside the sample, whereas here it is focused several microns below the sample.

\begin{figure}
\centering
\includegraphics[width=1\textwidth]{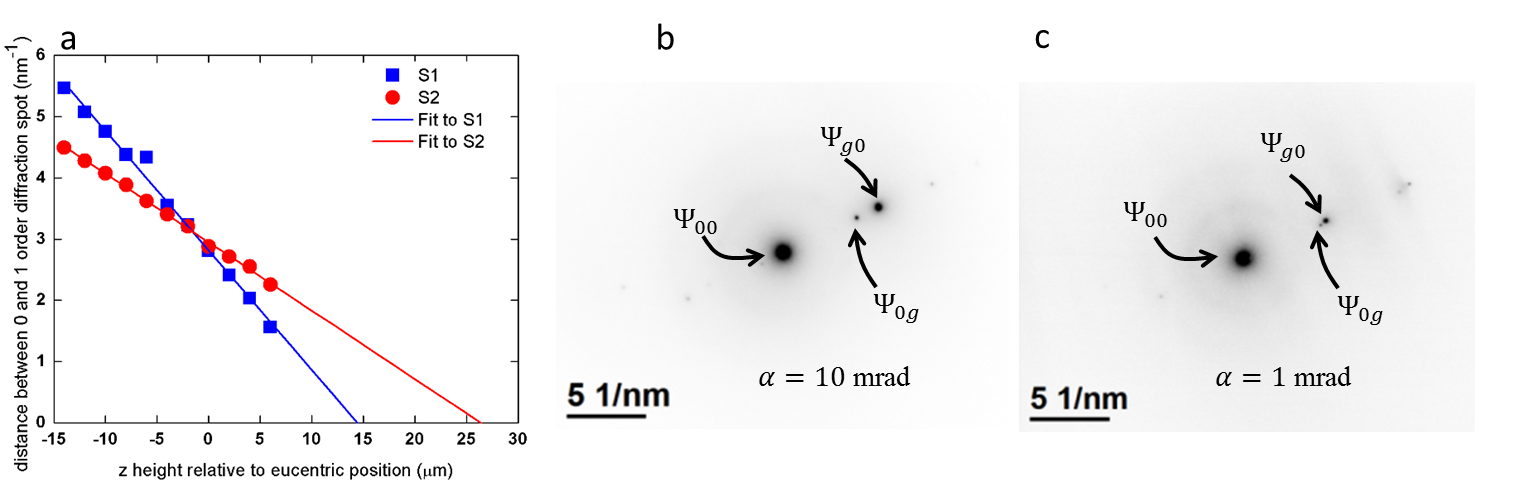}
\caption{Verifying the role of beam convergence in creating multiple diffraction spots. (a) Reciprocal space distance between primary and first order ($\textbf{g}=(1\bar{1}1)$)) diffracted spots from a single layer of silicon in CP2, as a function of the stage height $z$ relative to the eucentric plane, for convergence angles $\alpha_1=5$ mrad (red circles) and $\alpha_2=10$ mrad (blue squares). The solid lines are linear fits to the data points. The distance between the spots changed with the stage height for convergent beams. (b) At a large convergence semi-angle ($\alpha=10$ mrad), the diffraction spots from the $\Psi_{\textbf{0g}}$ and $\Psi_{\textbf{g0}}$ beams ($\textbf{g}=(0\bar{2}2)$)) at CP2 were distinct. (c) At a smaller  convergence semi-angle ($\alpha=1$ mrad), the two diffraction spots were much closer to each other. }
\label{ver}
\end{figure}

To verify this explanation of the observed diffraction pattern we performed the following experiments:
\begin{enumerate}
\item We first verified that there is significant horizontal displacement between the zero and first order diffracted beams at CP 2 from a single layer of silicon with a convergent beam. We fabricated a single grating sample and recording the diffraction pattern from it at CP 2 and different stage heights in the TEM. As can be seen from figure S\ref{ver}(a), this reciprocal space distance between the primary and first order diffracted spots decreases linearly with stage height. The red and blue data points depict two different beam semi-convergence angles $\alpha_{1} = 5$ mrad and $\alpha_{2} = 10$ mrad . This change in distance did not occur when imaging in the back focal plane with a parallel beam. This observation confirmed that the source of the change in distance was the convergence of the beam. As described earlier, the two-grating diffraction pattern can be thought of as a superposition of the patterns from a single layer at two heights separated by the inter-grating gap $D$ (with additional spots from double diffraction), and this experiment indicated that the spots from the two gratings were horizontally displaced due to the convergent of the beam. Also, by extrapolating the best fit curves we calculated that the convergent beams were focused $\sim$ 14 and \SI{26}{\micro\metre} below the eucentric plane for  $\alpha_{1}$ and $\alpha_{2}$ respectively. This plane was the position of CP 1 in this experiment.
\item We next placed the \SI{2.5}{\micro\metre} gap two-grating sample in the two-beam condition by tilting the TEM sample holder, so that we could focus on just one diffraction order($\textbf{g}_{\textrm{net}}=(0\bar{2}2)$)  besides the primary beam. We then varied the beam convergence angle (by changing the beam spot size) and recorded the DP at CP 2. Figure S\ref{ver}(b) and (c) show the recorded DP for $\alpha=$ of 10 and 1 mrad. The gap between the two ($0\bar{2}2$) spots arising from the beams $\Psi_{\textbf{0g}}$ and $\Psi_{\textbf{g0}}$ (with $\textbf{g}=(0\bar{2}2)$)) reduced at a lower convergence angle. This experiment verified that the gap between the diffracted spots from the two gratings could be tuned by changing $\alpha$, as expected from the proposed mechanism.
\end{enumerate}

\section*{Beam diameter and convergence angles for separation}
To determine the optimal beam diameter for simulations, we performed preliminary experiments with beam diameters ranging from 60-300 nm and $\alpha$ between 0.5 and 5 mrad. On the JEOL 2010F, beam diameters between 60-200 nm with $\alpha$ less than 4 mrad required the use of a very small (10 $\mu$m) condenser aperture which severely limited the intensity of the beams. The reduced beam intensity increased the exposure time required to record interference fringes. The increased exposure time resulted in poor fringe contrast due to stage vibrations. Thus, we chose a beam diameter greater than 200 nm and a relatively large $\alpha$ of 4 mrad. 

\begin{figure}
\includegraphics[width=1\textwidth]{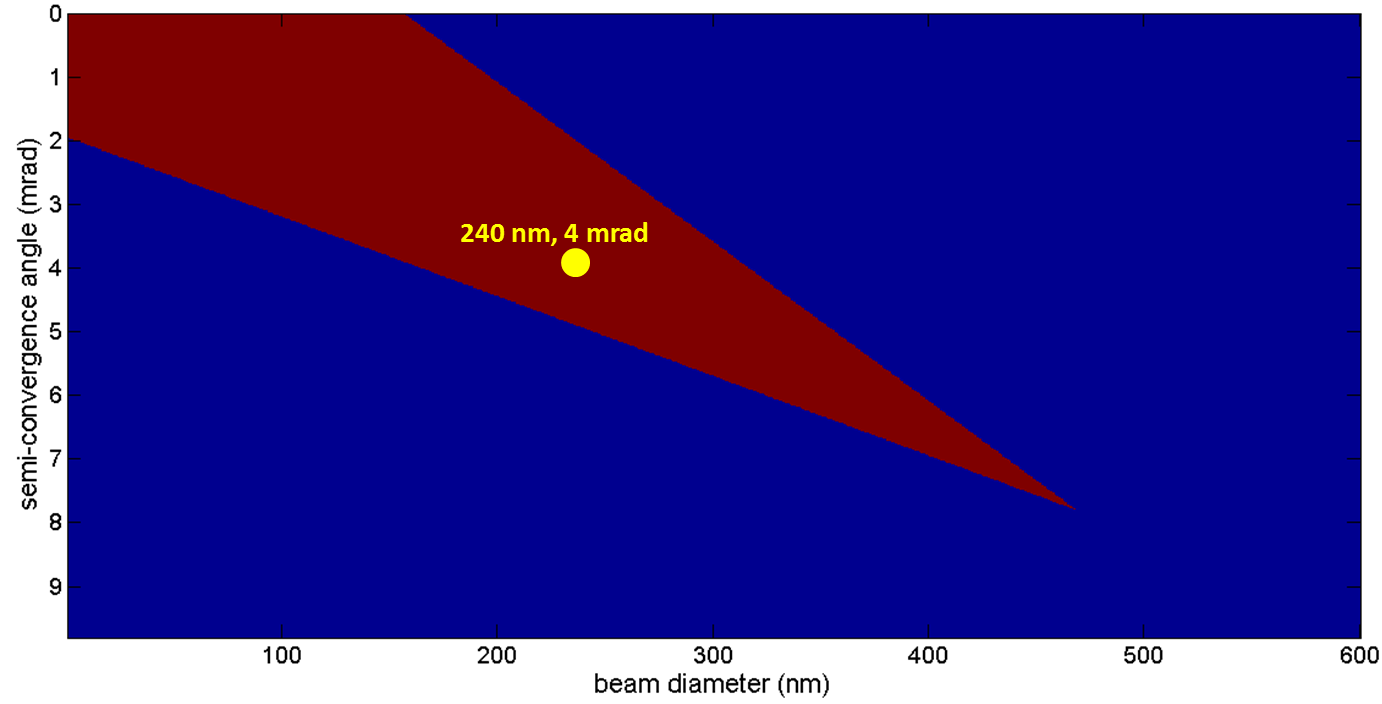}
\caption{Allowed values of beam semi-convergence angle and diameter at the first grating. The parameters for which results are reported in the paper are indicated by the yellow point.}
\label{para}
\end{figure}

Figure S\ref{para} depicts in red the range of beam diameters (on the first grating) and convergence angles that prevents the beams with $\textbf{g}_{\textrm{net}}=(000)$ and $\textbf{g}_{\textrm{net}}=(1\bar{1}1)$ from overlapping with each other at the second grating and the interference plane. The values in blue violate either one or both of these requirements. These values were obtained from the GSM simulations outlined in the paper, for a 200 kV beam and \SI{20}{\micro\metre} gap gratings. The interferometry results in the paper are for the beam parameters indicated by the yellow point (diameter 240 nm, $\alpha$ = 4 mrad). 

